\documentclass[useAMS,usegraphicx,usenatbib]{mn2e}
\usepackage[a4paper,centering, totalwidth=520pt, totalheight=700pt]{geometry}
\usepackage[position=top,labelformat=empty]{subfig}
\usepackage{graphicx}
\usepackage{aas_macros}     
\usepackage{amsmath}
\usepackage{amssymb}
\usepackage{fixltx2e}[1999/12/01]
\usepackage{tikz}
\usepackage{txfonts}

\newcommand{\Mass}{ \ensuremath{ h^{-1} M_{\odot}} }

\newcommand{\Mpc}{ \ensuremath{h^{-1} {\rm Mpc}} }
\newcommand{\Gpc}{ \ensuremath{h^{-1} {\rm Gpc}} }
\newcommand{\hMpc}{ \ensuremath{ h {\rm Mpc^{-1}}} }

\newcommand{\fap}{{paired-and-fixed }}
\newcommand{\dd}{\mathrm{d}}
\newcommand{\deltaNL}{\delta^{\mathrm{NL}}}
\newcommand{\deltaL}{\delta^{\mathrm{L}}}
\newcommand{\PL}{P^{\mathrm{\,L}}}

\newcommand{\hatPNL}{\hat{P}^{\mathrm{\,NL}}}
\newcommand{\hatPL}{\hat{P}^{\mathrm{\,L}}}
\renewcommand{\vec}[1]{\boldsymbol{\mathbf{#1}}}
\renewcommand{\Pr}{\mathrm{Pr}}

\def\simlt{\lower.5ex\hbox{$\; \buildrel < \over \sim \;$}}
\definecolor{darkgreen}{rgb}{0.0, 0.5, 0.0}

\title[Simulations with suppressed variance]{Cosmological $N$-body simulations with suppressed variance}

\begin{document}
\setlength{\topmargin}{-1.cm}

\author[Angulo \& Pontzen]{
\parbox[h]{\textwidth}
{Raul E. Angulo$^{1} \thanks{rangulo@cefca.es}$ \&
Andrew Pontzen$^{2} \thanks{a.pontzen@ucl.ac.uk}$}
\\
$^1$ Centro de Estudios de F\'isica del Cosmos de Arag\'on (CEFCA), Plaza San Juan 1, Planta-2, 44001, Spain. \\
$^2$ Department of Physics and Astronomy, University College London, Gower Street, London, WC1E 6BT, UK. \\
}
\maketitle

\date{\today}
\pagerange{\pageref{firstpage}--\pageref{lastpage}} \pubyear{2016}
\label{firstpage}

\begin{abstract} 
  We present and test a method that dramatically reduces variance
  arising from the sparse sampling of wavemodes in cosmological
  simulations. The method uses two simulations which are {\it fixed}
  (the initial Fourier mode amplitudes are fixed to the ensemble
  average power spectrum) and {\it paired} (with initial modes exactly
  out of phase). We measure the power spectrum, monopole and
  quadrupole redshift-space correlation functions, halo mass function
  and reduced bispectrum at $z=1$. By these measures, predictions from
  a fixed pair can be as precise on non-linear scales as an average over
  50 traditional simulations. The fixing procedure introduces a
  non-Gaussian correction to the initial conditions; we give an analytic argument
  showing why the simulations are still able to predict the mean properties of
  the Gaussian ensemble. We anticipate that the method will drive down the
  computational time requirements for accurate large-scale explorations of galaxy
  bias and clustering statistics, enabling more precise comparisons with
  theoretical models, and facilitating the use of numerical simulations in
  cosmological data interpretation.
\end{abstract}

\section{Introduction}

Numerical simulations are an essential tool for cosmology, especially for
interpreting observational surveys (see \citeauthor{Kuhlen2012} 2012 for a
review). They can be deployed to probe the impact of a given cosmological
ingredient \citep[e.g.][]{Baldi2014}, create virtual galaxy populations
\citep[e.g.][]{Overzier2009}, check and develop analytic treatments for
structure formation \citep[e.g.][]{Carlson2009}, and understand systematic and
statistical errors in cosmological measurements \citep[e.g.][]{Manera2015}.  In
the future, simulations could even be used to constrain cosmological parameters
\citep{AnguloHilbert2015}.

However, a limitation for all the above applications is the sparse sampling of
Fourier modes due to the finite extent of the simulation box. A given
cosmological simulation is initialised to a particular realisation of a
Gaussian random field. The power spectrum of the realisation, $\hatPL(k)$,
therefore differs from the ensemble mean power spectrum, $\PL(k)$. Given a
box large enough to capture all physical effects \citep{Bagla2009},
the largest-scale modes are still poorly sampled. This, together with
the non-linear coupling of small and large scales, implies
that several-Gpc size boxes generate statistical errors which limit inferences
on $100$ or even $10\,\mathrm{Mpc}$ scales.

This under-sampling effect is closely connected to (though, owing to
the non-linear evolution, not precisely the same as) observational
{\it cosmic variance}.  In the observational case, the finite volume
that can be achieved by a given survey constitutes an irreducible
source of uncertainty. On the other hand the computational variance
can be strongly suppressed, at least in principle, until it is smaller
than the cosmic variance and other sources of error. This is usually
achieved by simulating huge cosmological volumes
\citep[e.g.][]{Rasera2014} or a large number of realisations
\citep[e.g.][]{Takahashi2009}. Finite computing resources then generate a
tension between the need for large volumes and for high resolution
(the latter is required to better resolve the distribution of
individual galaxies and their internal structure). Even as
supercomputing facilities expand, the tension is becoming more acute
as surveys probe larger scales and constrain the statistics of
fluctuations to greater precision. For instance, reaching 1\% accuracy
over the whole range of scales to be probed by Euclid would require
the simulation of $\sim10^{5}$ Gpc$^3$.

In this {\it Letter} we propose and test a method to suppress the effect of box
variance drastically. We will show that with just two simulations we can
achieve the accuracy delivered by tens to hundreds of traditional simulations
at the same scale, depending on the particular problem in hand. Briefly, the
two simulations:

\begin{enumerate}
\item use a {\it fixed} input power spectrum, meaning that we enforce 
      $\hatPL = \PL$ when generating the initial conditions;
\item are {\it paired}, so that a hierarchy of effects due to chance 
      phase correlations can be cancelled \citep{Pontzen2015}.
\end{enumerate}

The first condition destroys the statistical Gaussianity of the input field
which, at first sight, would seem to limit the usefulness of the approach
\citep[see also][]{Neyrinck2013}.  However we will demonstrate empirically and
analytically that, by all measures explored here, the non-Gaussian corrections
have a negligible effect on ensemble mean clustering statistics.

This {\it Letter} is set out as follows. In Section \ref{sec:tests} we
implement and test our method. In particular, we quantify its
performance by comparing predictions with those from an ensemble of
300 independent simulations. We develop an analytic understanding of
why the method works in Section \ref{sec:method}. Finally, in Section
\ref{sec:conclusions} we present our conclusions.

\section{Comparison with an ensemble of simulations}
\label{sec:tests}

\subsection{Numerical Simulations}\label{sec:numerical-simulations}

\begin{figure}
\includegraphics[width=0.5\textwidth]{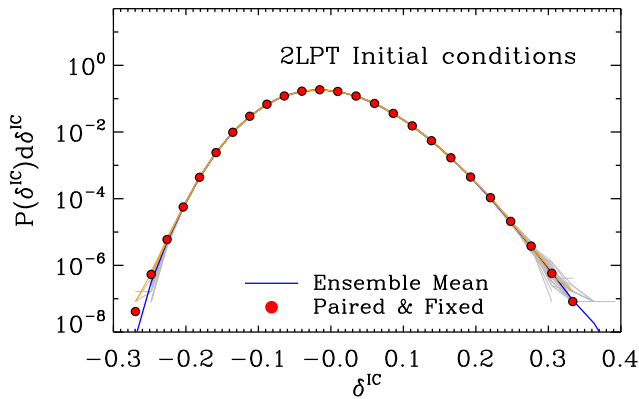}
\caption{The probability density function of non-linear overdensities in spheres
of radius $8\,\Mpc$, as measured in the fixed simulations (orange lines) and in
an ensemble of Gaussian simulations (grey lines), at the starting redshift,
$z=9$. The blue line and red circles show the mean of the respective cases, as indicated
by the legend.
\label{fig:pdf-1pt}}
\end{figure}

All simulations considered in this {\it Letter} contain $1024^3$
particles of mass $1.7\times10^{12}\,\Mass$ inside a box of side
$L=3\,\Gpc$. The initial particle positions are computed from
an input linear density field $\deltaL$ using 2LPT. The simulation
particles are then evolved under self-gravity with a COLA algorithm
\citep{Tassev2013} using 10 steps from $z=9$ to $z=1$. The
cosmological parameters assumed correspond to those of the Millennium
series \citep{Springel2005b}: 
$\Omega_m=0.25$, $\sigma_8=0.9$, and $h=0.73$.

The COLA algorithm is an approximate $N$-body method, in the sense
that the orbits inside high density regions are not properly
integrated.  However, the non-linear evolution of intermediate and
large scales is accurately captured \citep{Howlett2015,Koda2016}, at a
fraction of the computational cost of a traditional $N$-body
simulation. This enables the rapid simulation of extremely large
volumes, which in turn allows very precise calculations of different
statistics that serve as a benchmark for the performance of our
method. The total volume of our reference ensemble is
$8100\,h^{-1}\,\rm{Gpc}^{3}$; more details are given in
Chaves-Montero et al. (in prep).

The only difference between the ensemble of simulations and the pair
of fixed simulations is in the input fields $\deltaL(\vec{x})$. Because of
the finite box size, the Fourier modes for the field are quantised; we write

\begin{equation}
\deltaL(\vec{x}) \equiv \sum_i e^{i\vec{k}_i \cdot \vec{x}} \deltaL_i\textrm{,}
\end{equation}

\noindent where $i$ indexes the possible modes and $\deltaL_i$ is the Fourier
amplitude for the mode at wavevector $\vec{k}_i$. We can choose the indexing
such that $\vec{k}_{-i} = -\vec{k}_i$; note that, for the field
$\deltaL(\vec{x})$ to remain real, $\delta_i^{\mathrm{L}*} = \deltaL_{-i}$.

The reference ensemble of 300 simulations consists of boxes each with
$\deltaL_i$s drawn from a Gaussian, zero-mean probability distribution function
(pdf). Decomposed into the the magnitude $|\deltaL_i|$ and phase $\theta_i
\equiv \arg \deltaL_i $, the pdf for each independent mode $i$ is given by

\begin{equation}
 \Pr_g\left(\left|\deltaL_i\right|, \theta_i\right) \equiv 
  \frac{|\deltaL_i|}{\pi P_i }  \exp\left(-\frac{\left|\deltaL_i\right|^2}{P_i}\right), \label{eq:pdf-Gaussian}
\end{equation}

\noindent where $P_i$ is the discrete version of the power spectrum
$P(k)$.  
In the fixed-power approach, the pdf for mode $i$ is instead given by 
\begin{equation}
 \Pr_f\left(\left|\deltaL_i\right|, \theta_i\right) \equiv 
  \frac{1}{2 \pi} \delta_D\left(\left|\deltaL_i\right| - \sqrt{P_i}\right), \label{eq:pdf-fixed}
\end{equation}
where $\delta_D$ indicates the Dirac delta-function. One can sample from
$\Pr_f$ straightforwardly by setting
\begin{equation}
\deltaL_i = \sqrt{P_i} \exp \left( i \theta_i\right),\label{eq:delta-from-theta}
\end{equation}
with $\theta_i$ drawn with uniform probability between $0$ and $2 \pi$, and
$\theta_{-i} = - \theta_i$. The second of the pair of simulations is then
generated by transforming $\theta_i \to \pi + \theta_i$ \citep{Pontzen2015}.

Sampling from $\Pr_f$ results in an ensemble that is not equivalent to sampling
from $\Pr_g$. However, $\Pr_f$ can stand in place of $\Pr_g$ for many practical
calculations (the analytic justification is discussed in Section
\ref{sec:method}). We verified that, despite the fixed amplitudes, the
one-point input overdensity pdf in real space, $\deltaL(\vec{x})$, is still a
Gaussian deviate owing to the central limit theorem. Furthermore, Fig.
\ref{fig:pdf-1pt} shows the distribution of overdensities in the initial
conditions at $z=9$, $\delta^{IC}$, averaged over spheres of $8\,\Mpc$ radius
for a subset of traditional simulations (grey lines) and the two \fap (orange
lines; these overlap almost perfectly). The corresponding pdf for the combined
volume of the two paired-and-fixed simulations is shown by the red dots. There
is excellent agreement between this characterisation of the density fields of
traditional and fixed simulations, with both following a near-Gaussian
distribution. The mild skewness (which also agrees between the cases) arises
from the 2LPT particle displacements.

\subsection{Results}

\begin{figure}
\includegraphics[width=0.45\textwidth]{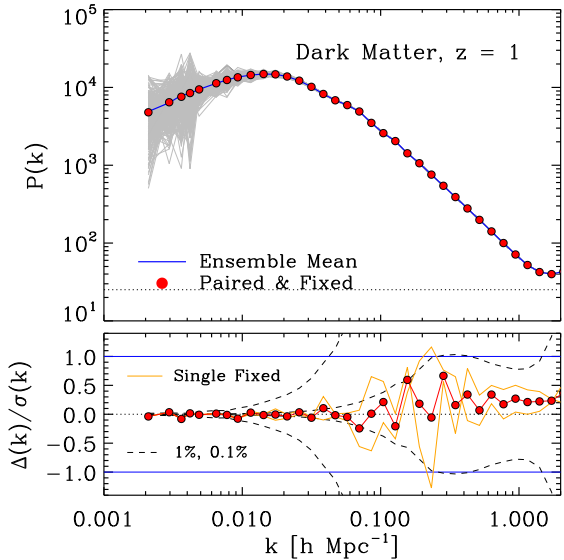}
\caption{The power spectrum of the dark matter at $z=1$. In the top panel,
measurements from the ensemble of 300 traditional simulations are shown as grey
lines, with the mean shown by a blue line. The solid 
red circles show the average of the two simulations in the \fap
set. Finally, the
horizontal dotted line marks the shot noise limit. In the bottom panel
we show the differences with respect to the average ensemble measurement, 
in units of the standard deviation in the ensemble. As in the top panel,
red symbols show the final estimate from the pair of fixed
simulations. We additionally show residuals in each of
the two individual fixed simulations by the orange lines. The 
envelopes bounded by dashed lines mark a $1\%$ (left) and $0.1\%$
(right) uncertainty in the power spectrum. The fixed pair
produces a power spectrum estimate with an r.m.s. error of just $0.27
\sigma$ on non-linear scales $0.03 < k / h \mathrm{Mpc}^{-1} < 1$. 
\label{fig:pk}}
\end{figure}

Fig. \ref{fig:pk} shows the dark matter power spectrum measured from the $z=1$
outputs. In the top panel, the results of the fixed pair (red circles) are
indistinguishable from the traditional ensemble mean (blue line) over all the
scales plotted, confirming that the approach correctly predicts the ensemble
average power spectrum in linear and in non-linear regimes. 

The bottom panel shows deviations with respect to the ensemble mean, in units
of the standard deviation of the ensemble, $\sigma(k) = (\langle \hatPNL(k)^2
\rangle - \langle \hatPNL(k) \rangle^2)^{1/2}$. On scales where evolution is
linear (approximately $k < 0.03\,\hMpc$) the fixed simulations should exactly
coincide with linear theory by construction. As expected, the measured power
spectrum agrees with the ensemble mean to an accuracy limited only by the
statistical errors of the latter, $\sigma(k)/\sqrt{300} \lesssim 2\%$ of
$P(k)$. At larger $k$, non-linear effects --- which depend not only on the
initial amplitude of Fourier modes but also on phases --- become important.
Accordingly, the power spectrum of the two individual fixed simulations (orange
lines) drift away from the exact mean. However the leading order deviations
from the ensemble mean are equal and opposite in sign \citep{Pontzen2015}
between the pair of fixed simulations, so that their average (red dots) has a
reduced r.m.s. error much below $1\%$ ($0.27\sigma$ over the range $0.03 <
k\,\hMpc < 1$). The accuracy of our pair of fixed simulations by this measure
is approximately equivalent to averaging over 14 traditional simulations,
allowing for a factor 7 reduction in computer time. In particular, the
technique suppresses statistical errors on {\it all} scales to the point where
they are smaller than the impact of numerical parameters \citep{Schneider2015}.

\begin{figure}
\includegraphics[width=0.45\textwidth]{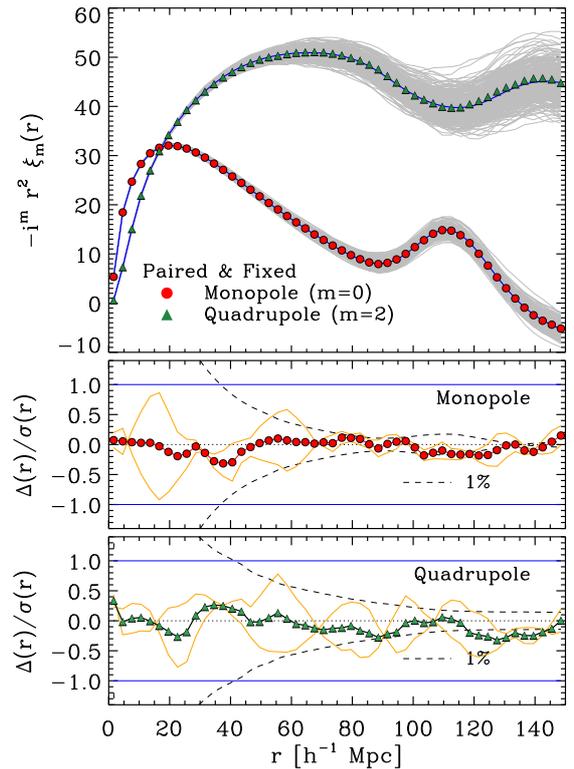}
\caption{ Same as Fig. \ref{fig:pk} but for the monopole (red circles) and the
quadrupole (green triangles) of the redshift-space correlation function. The
r.m.s. error on the \fap result is $0.12\,\sigma$ and $0.17\,\sigma$ for the
monopole and quadrupole respectively, meaning that around 50 traditional
simulations are required to reach the accuracy of a fixed pair of simulations.
\label{fig:xi}}
\end{figure}

In Fig. \ref{fig:xi} we show that the high accuracy of the method also holds in
redshift space. In this figure we plot the monopole (red circles) and
quadrupole (green triangles) terms of an expansion of the 2D correlation
function in terms of Legendre polynomials. Predictions from the pair of fixed
simulations again agree well with the ensemble mean. The same pattern persists
where the individual fixed simulations perform best on large scales, while on
smaller scales the pairing leads to a substantial cancellation of remaining
errors. The overall technique yields a precise prediction for the non-linear
correlation function, reaching a $2\%$ accuracy over the whole range of scales
investigated (in particular around the baryonic acoustic oscillation peak,
whose shape and location is currently driving large simulation campaigns). With
traditional ensemble-average techniques, achieving this accuracy would require
around $50$ simulations of $3\,\Gpc$ box size.
 
\begin{figure}
\includegraphics[width=0.45\textwidth]{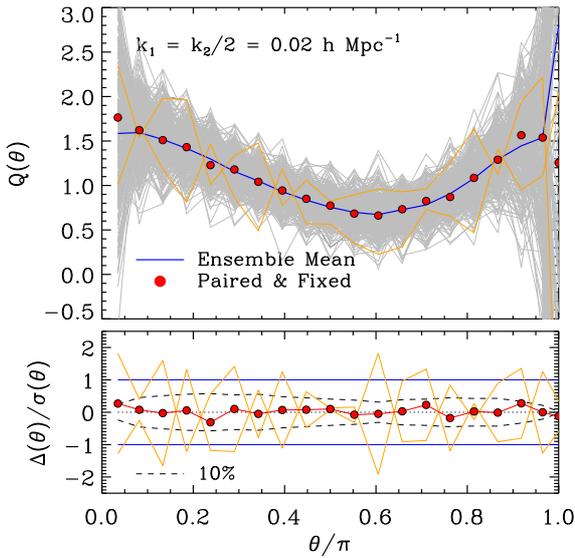}
\caption{Same as Fig. \ref{fig:pk} but for the reduced bispectrum. The 
configuration plotted corresponds to triangles with two sides fixed
at $k_1 = 0.02\,\hMpc$ and $k_2 = 0.04\,\hMpc$, with their angle ranging
from 0 to $\pi$. The r.m.s. deviation is $0.14\,\sigma$.
\label{fig:bis}}
\end{figure}

Having established the accuracy of our simulations for predicting two-point
statistics, we now turn to higher-order clustering. The bispectrum is
defined (in the limit that the box size is infinite) by
\begin{equation}
B(k_1,k_2,\theta)\, \delta_D(\vec{k}_1 + \vec{k}_2 + \vec{k}_3) = \langle \deltaNL(\vec{k}_1) \deltaNL(\vec{k}_2)
\delta^{\mathrm{NL}}(\vec{k}_3) \rangle\textrm{,}
\end{equation}

\noindent where $\deltaNL(\vec{k})$ is the Fourier transform of the non-linear
evolved overdensity. We have assumed statistical isotropy in writing $B$ as a
function of $\theta$, the angle between the $\vec{k_1}$ and $\vec{k_2}$
vectors, and statistical homogeneity imposes the Dirac-delta dependence on the
left-hand-side. We particularly consider the case where $k_1 = 0.02\,\Mpc$ and
$k_2 = 0.04\,\Mpc$ to capture the onset of non-linearity, and plot the reduced
bispectrum
\begin{equation}
Q(\theta) = \frac{\hat{B}(k_1,k_2,\theta)}{ \hatPNL(k_1)\hatPNL(k_2) + \hatPNL(k_1)\hatPNL(k_3) + \hatPNL(k_2)\hatPNL(k_3)},
\end{equation}

\noindent where $\hat{B}$ is the estimated bispectrum from a simulation. The
definition of $Q(\theta)$ divides out much of the sensitivity to the power
spectrum realisation. Accordingly, when we plot this quantity in
Fig.~\ref{fig:bis}, each of the two individual fixed simulations exhibit
fluctuations of an amplitude comparable to that in traditional realisations.
However, the pairing procedure cancels the leading-order contribution to these
fluctuations because they have odd parity in the input linear density field.
Therefore the final estimate from the pair of fixed simulations has an r.m.s.
deviation from the ensemble average of only $0.14\sigma$ over all $\theta$.
Reaching this accuracy with traditional simulations would again require
averaging over 50 (as opposed to two) realisations.

\begin{figure}
\includegraphics[width=0.45\textwidth]{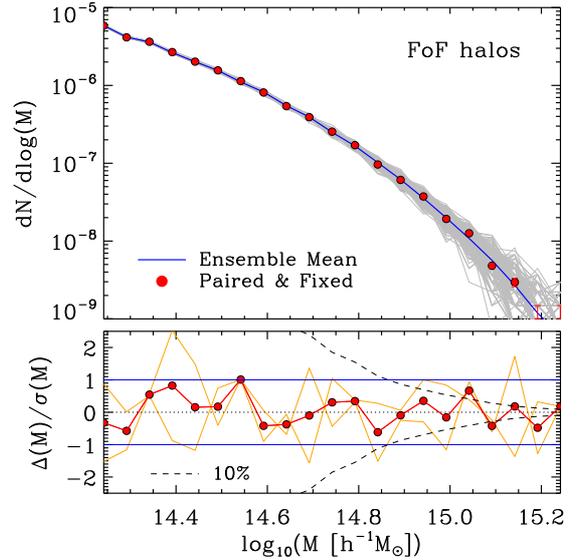}
\caption{Same as Fig. \ref{fig:pk} but for the abundance of FoF halos. The
r.m.s. deviation is $0.47\,\sigma$ -- this is a significant improvement over
two randomly chosen ensemble members ($0.70 \sigma$) albeit not as decisive as
in the earlier cases of the power spectrum, correlation function and
bispectrum. \label{fig:mf}}
\end{figure}

As discussed in \cite{Koda2016}, the COLA $N$-body algorithm does not resolve
the internal structure of halos but nonetheless predicts accurate mass
functions for the overall population. Therefore we can meaningfully test the
abundance of collapsed objects. In Fig. \ref{fig:mf} we show the mass function
of dark matter halos identified using a Friends-of-Friends algorithm
\citep{Davis1985} with a linking length set to $l=0.2$. We find the same
qualitative picture as in previous plots and statistics, although with slightly
less striking noise suppression. The new method produces results with
suppressed fluctuations relative to two Gaussian simulations, with strong
cancellations between the pair. The average r.m.s. error is $0.47\sigma$,
roughly the level expected from four simulations randomly picked from the
traditional ensemble.

\section{Analytic exploration}
\label{sec:method}

In the previous Section we showed that \fap simulations are able to predict the
average properties of a traditional ensemble.  We will now explore
the technique from an analytic perspective.  The pairing approach has recently been
introduced and discussed elsewhere \citep[][see particularly section
II.C]{Pontzen2015} and so our focus is on the power spectrum fixing.
Sampling from the $\hat{P}(k)$-fixed pdf $\Pr_f$, defined by
equation~\eqref{eq:pdf-fixed}, is not equivalent to sampling from the
true Gaussian $\Pr_g$, equation~\eqref{eq:pdf-Gaussian}. The aim of
this Section is therefore to motivate more precisely why $\Pr_f$
has reproduced the ensemble average results of $\Pr_g$.

Expectation values of any $n$-point expression with respect to
either $\Pr_f$ or $\Pr_g$ will be denoted by $\langle \delta_{i_1}
\cdots \delta_{i_n} \rangle_f$ and $\langle \delta_{i_1} \cdots
\delta_{i_n} \rangle_g$ respectively. In the case of the fixed
distribution, we can use expression~(\ref{eq:delta-from-theta}) to
write that
\begin{equation}
\langle \deltaL_{i_1} \cdots \deltaL_{i_n} \rangle_f = \frac{\sqrt{P_{i_1} \cdots P_{i_n}}}{(2 \pi)^N}\int_0^{2 \pi}
\dd^N \theta \, \exp\left(i\theta_{i_1} + \cdots + i\theta_{i_n} \right)\textrm{,}\label{eq:correlation-fixed}
\end{equation}
\noindent where the integral is over the possible $\theta$ values for all $N$ modes.

For $n=1$ the single phase factor
$\exp (i \theta_1)$ averages to zero, and consequently $\langle \deltaL_i \rangle_g =
\langle \deltaL_i \rangle_f = 0$.  This result extends to any $n$-point correlation for $n$ odd; we therefore need only consider
the even-$n$ cases further.

For $n=2$, the properties of the two pdfs are indistinguishable: 
\begin{equation}
\langle \deltaL_i \deltaL_j \rangle_f = \langle \deltaL_i \deltaL_j \rangle_g = \delta_{i,-j}
P_i\textrm{,}
\end{equation}
where $\delta_{i,-j}$ is the Kronecker delta equal to $1$ when $i=-j$
and $0$ otherwise, and there is no sum
implied over repeated indices. The Gaussian result is standard, and
the fixed result is obtained by seeing that when $i \ne -j$, the $i$
and $j$ phase integrals in equation (\ref{eq:correlation-fixed})
evaluate to zero. For $n=4$, the Gaussian result follows by Wick's
theorem:
\begin{equation}
\langle \deltaL_i \deltaL_j \deltaL_k \deltaL_l \rangle_g =
\delta_{i,-j}\delta_{k,-l} P_i P_k + \delta_{i,-k} \delta_{j,-l} P_i P_j +
\delta_{i,-l} \delta_{j,-k} P_i P_j\textrm{.} \label{eq:wick-4}
\end{equation}
The fixed result, again obtained through use of~\eqref{eq:correlation-fixed} is
similar to the Gaussian case because indices must be ``paired up'' for their
phase integrals to be non-vanishing. The only difference arises in
the case where $\delta_i \delta_j \delta_k \delta_l = |\delta_i|^4$; here, the Gaussian result is
$\langle |\deltaL_i|^4 \rangle_g=3 P_i^2$ but in the fixed case we find that
$\langle |\deltaL_i|^4 \rangle_f = P_i^2$. Overall the result is~therefore
\begin{equation}
  \langle \deltaL_i \deltaL_j \deltaL_k \deltaL_l \rangle_f =
  \langle \deltaL_i \deltaL_j \deltaL_k \deltaL_l \rangle_g - 2\left(\delta_{ij} \delta_{kl}
    \delta_{i,-k} + \delta_{ik} \delta_{jl} \delta_{i,-j} + \delta_{il}\delta_{jk}\delta_{i,-j} \right) P_i^2 \textrm{,} \label{eq:gaussian-to-fixed-4}
\end{equation}
\noindent assuming that we have $P_0=0$ (otherwise a further term is
necessary to divide the correction by $3$ in the $i=j=k=l=0$ case).  

The correction~\eqref{eq:gaussian-to-fixed-4} is consistent with how the power
spectrum of a fixed realisation must have zero variance:
\begin{equation}
\left\langle \left(|\deltaL_i|^2 - P_i\right)^2 \right\rangle_f =
\left\langle \left(|\deltaL_i|^2 - P_i\right)^2 \right\rangle_g - 2
P_i^2 = 0\textrm{.}\label{eq:zero-variance}
\end{equation}
Evidently there is a dramatic difference --- intentionally so ---
between fixed and Gaussian statistics: in the linear regime, the fixed
$P(k)$ approach reproduces the ensemble mean with no variance.  While
this also means that the input trispectrum is unavoidably non-Gaussian
by equation \eqref{eq:gaussian-to-fixed-4}, the correction only
appears when all indices always take the same value (up to sign). We can now
explain why most measures of the output non-linear density field are
extremely insensitive to this change.

The non-linear density field can be written in standard perturbation theory
\citep[SPT, e.g.][]{Bernardeau2002} as

\begin{equation}
\deltaNL_i = \deltaL_i + \sum_{jk} F^{(2)}_{ijk} \deltaL_j \deltaL_k +
\sum_{ijkl} F^{(3)}_{ijkl} \deltaL_j \deltaL_k \deltaL_l + \cdots
\end{equation}

\noindent where $F^{(n)}_{\cdots}$ for $n=2, 3, \cdots$ are the discretised
version of the SPT kernels which in turn are homogeneous, degree-zero,
continuous functions of the wavevectors. As a concrete example of an
observable correlation in this formalism, we can consider the one-loop
SPT non-linear power spectrum with Gaussian statistics:
\begin{align}
P^{\mathrm{NL,g}}_i & \equiv \langle \deltaNL_i \deltaNL_{-i}
                      \rangle_g \nonumber  \simeq P_i + \sum_{jklm} \left( F^{(2)}_{ijk} F^{(2)}_{-i,lm} + 2 \delta_{i,-j}
  F^{(3)}_{iklm} \right) \times \nonumber \\
& \left(\delta_{j,-k}\delta_{l,-m} P_j P_l +
  \delta_{j,-l} \delta_{k,-m} P_j P_k +
\delta_{j,-m} \delta_{k,-l} P_j P_k\right)\textrm{.}\label{eq:PNL-one-loop}
\end{align}
Momentum conservation implicit in the $F^{(n)}_{\cdots}$s and explicit
in the Kronecker deltas eliminate three of the summations, so that the
overall summation is over just one index. Therefore the magnitude of
the one-loop terms scales proportionally to $N_k P(k)$ where $k$
is a characteristic scale and $N_k$ is the number of modes around that
scale (as defined by the range of modes over which the relevant $F$ is
large).  In a continuum limit (i.e. as the box size $L \to \infty$),
$N_k$ turns into the appropriate Fourier-space volume. These simple
scaling behaviours are assured by the degree-zero homogeneity of the
$F^{(n)}$ functions.

In the fixed case, expression~\eqref{eq:PNL-one-loop}~must be
corrected by using relation~\eqref{eq:gaussian-to-fixed-4}, giving
\begin{align}
P^{\mathrm{NL,f}}_i \simeq P^{\mathrm{NL,g}}_i  - 12 F^{(3)}_{i,-i,i,-i}
  P_i^2 - 2 \sum_j  F^{(2)}_{ijj}
  F^{(2)}_{-i,-j,-j} P_j^2\textrm{,}\label{eq:PNL-one-loop-correction}
\end{align}
\noindent which is valid at one-loop order for the case $\vec{k}_i \ne 0$. Here
most of the Kronecker deltas have already been summed out; the remaining
summation, by momentum conservation in $F^{(2)}$, only has a contribution at
the index $j$ with $\vec{k}_j = \vec{k}_i/2$.  The overall contribution of the
correction~\eqref{eq:PNL-one-loop-correction} is therefore suppressed relative
to the physical terms in equation~\eqref{eq:PNL-one-loop} by
$\mathcal{O}(N_k)$.
 
For the bispectrum with Gaussian statistics, we have

\begin{align}
B^g_{ijk} & \equiv \langle \deltaNL_i \deltaNL_j \deltaNL_k \rangle_g
            \nonumber \\
& \simeq 2 F^{(2)}_{i,-j,-k} P_j P_k + \delta_{j,-k} \sum_l
            F^{(2)}_{il,-l} P_j P_l + \textrm{cyc. perms in } ijk 
\end{align}

\noindent to one-loop order. The second term contributes only for
$\vec{k}_i=0$. The correction is now
\begin{equation}
B^f_{ijk} = B^g_{ijk} - 4 F^{(2)}_{i,j,-j} P_j^2 \delta_{jk} - 2 F^{(2)}_{i,-j,-j} P_j^2 \delta_{jk}  -
\textrm{cyc. perms in } ijk\textrm{,}
\end{equation}
and is non-zero only in the case where
$\vec{k}_j = \vec{k}_k = -\vec{k}_i/2$ or $\vec{k}_i=0$ (or a cyclic
permutation of those configurations). All other bispectra
are unaffected by the changed statistics at this order.

For higher order perturbation theory (or higher-$n$ correlations) the
overall pattern established here will remain: the linear $n$-point
correction term \eqref{eq:gaussian-to-fixed-4} will always involve at
least one extra Kronecker delta relative to the physical part
\eqref{eq:wick-4}. For observable correlations, this implies that
either the effect is diluted by a power of a large factor $N_k$ (as in the case
of the one-loop power spectrum) or plays a role only in a measure-zero
part of the continuous function being studied (as in the case of the
one-loop bispectrum). The intentionally-reduced power spectrum variance
\eqref{eq:zero-variance} falls into the latter category, since it is
only the diagonal part of the full covariance that is
altered. 

\section{Conclusions}
\label{sec:conclusions}

In this {\it Letter} we have explored a new method to suppress the impact of
under-sampling Fourier modes in simulations. 

By fixing the initial amplitude of Fourier modes to the ensemble mean, variance
has been eliminated on linear scales. In the non-linear regime, the suppression
is imperfect because phase-correlation effects begin to impact on the evolved
amplitudes. However by also pairing the simulation with a phase-reversed
counterpart \citep{Pontzen2015} we can average away the leading-order
imperfections of this type; fluctuations about the ensemble mean are then
near-identical in magnitude but opposite in sign. Residual noise could be
reduced arbitrarily by considering an ensemble of \fap simulations with
different realisations of the initial random phases. 

We have tested the non-linear dark matter power spectrum, the
  multipoles of the redshift-space correlation function, the reduced
  bispectrum and the halo mass function. In all cases, the method is
  unbiased (up to the accuracy of our comparison ensemble averages)
  and strongly suppresses unwanted variance. These tests were carried
  out with a suite of 300 COLA simulations at $z=1$. The analytic
  arguments of Section \ref{sec:method} suggest that the accuracy of
  the results should be maintained to all redshifts. Similarly, we do
  not expect results to change when using more accurate integration
  methods than COLA, especially since small-scale gravitational
  collapse are largely insensitive to large-scale correlations. All
  these points deserve systematic investigation in future.

Paired simulations can be used with purely Gaussian initial conditions if
desired, retaining many of the small-scale benefits we have discussed.
Conversely, single unpaired simulations with fixed amplitudes can be used,
retaining the large-scale benefits. Whenever fixing is applied, the ensemble
statistics are not strictly Gaussian. The local one-point pdf is, however,
unaffected (Figure \ref{fig:pdf-1pt}) and furthermore our numerical results
directly show that a variety of statistics attain the correct, unbiased
ensemble mean value. We gave an analytic discussion of why the non-Gaussianity
does not impinge, arguing that the errors are either strongly suppressed by the
large density of modes or, in other cases, affect only a measure-zero set of
correlations. Fixing the power spectrum does need to be approached with care
but our results underline that it can be a valuable technique.

Straightforward applications are in any comparison to analytic models,
in characterisation of the performance of data modelling, in emulators
and in development of fitting functions for non-linear statistics. It will
be particularly valuable to couple the technique to high-resolution
simulations incorporating baryonic effects to measure galaxy bias, free
of the usual difficulties of large-scale variance. Furthermore the
method could be used in combination with rescaling
techniques to quickly predict galaxy clustering statistics as a
function of cosmological parameters \citep{Angulo2010}. All these are
crucial steps towards a comprehensive exploitation of upcoming
survey data.

\vspace{-0.5cm}

\section*{Acknowledgements}

We would like to thank Jon\'as Chavez-Montero for providing us with access to
the ensemble of COLA simulations. We thank the Lorentz Center and the
organisers of the ``Computational Cosmology'' workshop where this study was
initiated, and Oliver Hahn, Carlos Hernandez-Monteagudo, Aseem Paranjape,
Hiranya Peiris, An\v{z}e Slosar, and Matteo Viel for helpful discussions. REA
acknowledges support from AYA2015-66211-C2-2.  AP is supported by the Royal
Society.

\vspace{-0.5cm}

\bibliographystyle{mn2e} \bibliography{database}

\label{lastpage} \end{document}